\documentclass[twocolumn, showpacs,preprintnumbers,amsmath,amssymb]{revtex4}
\usepackage{amssymb}
\usepackage{xcolor}
\usepackage{graphicx}
\usepackage{dcolumn}
\usepackage{bm}
\usepackage{multirow}
\usepackage{booktabs}
\usepackage{natbib}
\usepackage{soul}
\usepackage{amssymb}
\usepackage{amsmath}	
\usepackage{color}
\usepackage{cases}

\begin{document}
\preprint{APS/123-QED}

\title{Study of bound and resonant states of NS molecule in the R-matrix approach}

\author{F. Iacob$^{1}$}\email[]{felix.iacob@gmail.com}
\author{Thomas Meltzer$^{2}$} 
\author{J. Zs Mezei$^{3,4}$}\email[]{mezei.zsolt@atomki.hu}
\author{I. F. Schneider$^{4,5}$}\email[]{ioan.schneider@univ-lehavre.fr}
\author{J. Tennyson$^{2,3,4,6}$}\email[]{j.tennyson@ucl.ac.uk}
\affiliation{$^{1}$Physics Faculty, West University of Timisoara, 300223,Timisoara, Romania}%
\affiliation{$^{2}$Department of Physics and Astronomy, University College London, WC1E 6BT London, UK}%
\affiliation{$^{3}$HUN-REN Institute for Nuclear Research (ATOMKI), H-4001 Debrecen, Hungary}%
\affiliation{$^{4}$LOMC-UMR6294, CNRS, Universit\'e Le Havre Normandie, 76600 Le Havre, France}%
\affiliation{$^{5}$LAC-UMR9188, CNRS Universit\'e Paris-Saclay, F-91405 Orsay, France}%
\affiliation{$^{6}$HAS-Distinguished Guest Scientists Fellowship Programme-2022}%
\date{\today}

\begin{abstract}
The bound and resonance states along with corresponding autoionization widths for nitrogen sulphide (NS) molecule are determined using electron NS$^+$ cation scattering calculations.  The calculations are performed for  $^2\Sigma^+$, $^2\Pi$ and $^2\Delta$ total symmetries using
the {\em ab initio} R-matrix method for both bound and continuum states. Calculations are performed on a grid of 106 points for internuclear separations between 1.32 and 3 \AA. 
The resonance states yield dissociative potential curves which, when considered together with their widths, provide input for models of different electron-cation collision processes including dissociative recombination, and rotational and vibrational excitation. Curves and couplings which will lead directly to dissociative recombination are identified.
\end{abstract}


\maketitle

	\section{Introduction}\label{intro} 
	
	Sulphur is the tenth and nitrogen is the fifth most abundant element in the Universe, thus their chemistry is of key importance for astronomical environments.
	It is also known to play a significant role in planetary geochemistry and polymer chemistry. 
	 NS emissions are accessible to astronomical observation and NS was among the first diatomic molecules to be observed in space \cite{Somerville}.
	The start of nitrogen sulphide (or more precisely mononitrogen monosulphide) chemistry, dates from the discovery of a compound named as   ``thiazyl'' radical $(\mbox{S}=\mbox{N}\cdot)$ by Demarçay \cite{Demarcay}. It was identified inside a polymeric structure [SN]$_x$ known as polythiazyl. Later it was proved that this non-metallic compound has superconducting properties at low temperatures and the study of this molecule began to command considerable attention \cite{Preuss}.  
	
	The situation was not the same for NS$^+$ cation. Although it is now known to be ubiquitous in the interstellar medium (ISM), it was only recently detected \cite{NS+disc}. 
	A year later, the first detection of NS$^+$  in a photodissociation region of the  Horsehead Nebula was reported \cite{Riviere19}. 
	
	A challenge for astrochemistry is to understand the mechanisms and rates of formation and destruction of both neutral and cationic molecules. Electron collisions with cations are a common mechanism that leads to their dissociation. 
    However, despite its astrochemical importance, no estimate of the dissociative recombination (DR) rate	of  NS$^+$ is available \cite{NS+disc}.
	Modelling this mechanism requires the calculation of the doubly excited states of the compound neutral system.

	To our knowledge, there are no previous studies of e + NS$^+$ scattering and relatively few 
	theoretical studies on the NS$^+$ target. These studies mostly concern NS electronic ground state. 
	O'Hare \cite{OHare70} performed a self-consistent field (SCF) calculation at the equilibrium geometry, R$_e$=1.4957~\AA, and estimated a dissociation energy of $D_0$=4.8 eV;
	other studies of equilibrium geometry of the X $^1\Sigma^+$ state of NS$^+$ can be found in \cite{Kapf} and reference within. 
	The low-lying valence states of NS$^+$ were explored by Karna {\it et al.} \cite{Karna86} using configuration interaction (CI) calculations. Most important for this work is the 
	more recent study of NS$^+$ metastable states  by Yaghlane and Hochlaf \cite{Ben09}.

	The purpose of this article is to provide reliable data which can be used as input to models
	capable of giving good estimates of the DR rate as well as rates for other important processes
	such as electron-impact vibrational excitation. 
	Our study of electron scattering from the NS$^+$ cation uses the R-matrix method to give 
	{\em ab initio} estimates of the
	bound and continuum states of NS. 
	We identify Feshbach resonances that can provide a route to DR by using a model comprising the ionic core in its ground $X\,^1\Sigma^+$ state and three lowest excited states, a$\,^3\Sigma^+$, b$\,^3\Pi$ and c$\,^3\Delta$. 
	Besides  potential energy curves (PEC) of the resonances, we provide their autoionization widths of these
	doubly excited states; we also provide PECs for bound, singly excited Rydberg states of NS.
	These Rydberg states are significantly more excited than those computed in the previous study
	of electronically excited states of NS \cite{Lie85}, where only $^2\Sigma^+$ symmetry states were considered. 
	These highly excited Rydberg states of neutral NS  provide an important component of the  DR mechanism for low-energy electron collisions with cations such as those encountered in the ISM.
	
	The article starts with a section detailing our calculations which considers the method,  the target model, and how our scattering calculations are performed. Section three presents and discusses
	our results which comprise bound states, resonances, widths and effective principal quantum numbers.
	Finally, we summarise the main results of this article in the conclusion with pointers towards future work. 

	\section{Calculations}
	
	\subsection{Method}
	The R-matrix method is widely used for studies of scattering problems \cite{Burke}. It was
	originally formulated for nuclear physics and given an {\it ab initio} variational formulation 
	for electron collision problems in atomic and molecular physics. A detailed discussion of its application to electron-molecule problems can be found in the extensive review \cite{tenn}.
	 
	Our calculations use the UKRmol+ code \cite{UKRmol+}, a new implementation of the time-independent UK R-matrix electron–molecule scattering code, which uses
	the electronic structure code MOLPRO \cite{molpro} to provide target
	wave functions. 
	
	The R-matrix method is based on the division of space into an inner and an outer region. 
	The inner region is a sphere,
	here with  radius of 10 a$ _0 $, centred on the centre-of-mass of the NS$^+$ target. 
	In the inner region we perform two separate calculations: 
	a target one which considers the $N$ interacting target electrons, 
	and an $N + 1$ electron calculation in which the scattering electron interacts with all target electrons.
	Then these two calculations are used in turn to provide R-matrices on the boundary which are then used in the outer region to calculate scattering properties.
	
	In the inner region the $(N+1)$-electron system of the target and the colliding electron are described by the wave function:
	\begin{eqnarray}\label{R-Mwf}\nonumber
	\Psi_k(&x_1&,\cdots,x_{N+1}) = \mathcal{A} \sum_{i,j} a_{i,j,k}\, \Phi_i^N (x_1,\cdots, x_N)\times\\
	&& u_{i,j}(N+1)+\sum_i b_{i,k}\, \chi_i^{N+1}(x_1,\cdots,x_{N+1}).
	\end{eqnarray}
	Here $\Phi_i^N$ are the wave functions describing the $i^{\rm th}$ target state. The $u_{i,j}$ represent the extra continuum orbitals. $\Phi_i^N$  and $\chi_i$ are constructed to vanish on the R-matrix boundary. Hence, the $\chi_i$, which represent extra configurations
	obtained by placing the scattering electron in a target orbital, are known as $L^2$ configurations.
	The wave function of the $N+1$ electron system has to obey the Pauli principle, which is achieved by the action of $\mathcal{A}$ which represents an anti-symmetrization operator.   The coefficients
	 $a_{i,j,k}$ and $b_{i,k}$ are variational coefficients of expansions~\cite{morgan}. 
	To construct the R-matrix on the boundary we used all the solutions of the inner region problem.
	
	Within the framework of the R-matrix method, there are a variety of different scattering models and procedures that can be used \cite{tenn}. Close-coupling (CC) expansions are used here which involve including several target states in Eq. (\ref{R-Mwf}).  The use of the CC method is essential for describing electronic excitation and is also best for studying Feshbach resonances, which is the major goal of this paper.
	
	\subsection{NS$^+$ target} \label{target-ss}
	
	A good description of the target is required when calculating the resonant scattering states of a molecule. A state averaged multi-configuration self-consistent field (MCSCF) model was used to generate the first four states of NS$^+$. These were computed with MOLPRO using a Gaussian-type orbital (GTO) cc-pVQZ basis set.  
	Several speed and accuracy tests were performed with different bases to decide the optimal one.
	The ground state of NS$^+$ is $X$ $^1\Sigma^+$ and the equilibrium bond length was determined at $1.44\pm 0.01$~\r{A}. 
	\begin{figure}
		  \includegraphics[width=1.15\columnwidth]{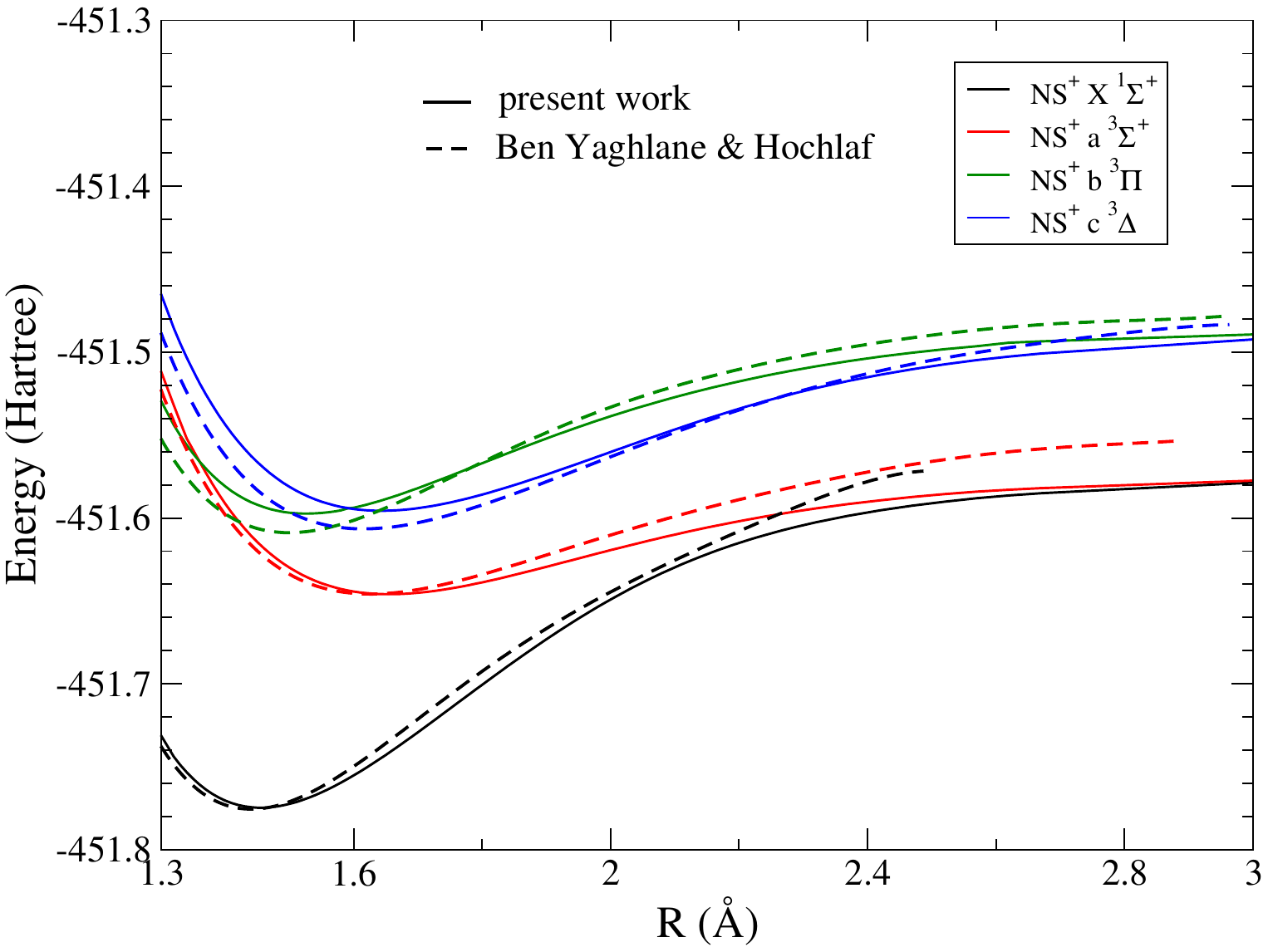}
		\caption{Potential energy curves for the ground ($X^1\Sigma^+$) and three lowest excited states (a$\,^3\Sigma^+$, b$\,^3\Pi$, c$\,^3\Delta$) of NS$^+$ target. 
			Comparison is given with the work of Ben Yaghlane and Hochlaf  \cite{Ben09}
		}
		\label{pes}
	\end{figure}
	The low-lying excited states are all triplets: a$\,^3\Sigma^+$, b$\,^3\Pi^+$, and c$\,^3\Delta^+$.
	Whilst NS$^+$ belongs to the C$_{\infty v}$ point group, both MOLPRO and UKRmol+ only allow
	the calculation to be performed in C$_{2v}$ symmetry. In this case 
	the molecular orbitals are labeled according to their symmetry properties as belonging to one of the four irreducible representations (A$_1$, A$_2$, B$_1$, and B$_2$). In the case of a linear
	molecule in the $C_{2v}$ point group, results obtained in the B$_1$ and B$_2$ representations are degenerate. 
	The 22 electrons of NS$^+$ are organized in (10,4,4,1)  orbitals of which: (4,1,1,0) are frozen, (4,2,2,0) are used for the complete active space (CAS) and (2,1,1,1) are used as
	virtual orbitals in the scattering calculation. 
	We note that the fewer frozen orbitals we consider and/or the larger the CAS, the higher accuracy we get but, on the other hand, this high accuracy requires more computer time.
	
	The NS$^+$ target potential energy curves we obtained are plotted in figure \ref{pes}.
	Our calculation is displayed using solid lines and compared with the  calculations by Ben Yaghlane and Hochlaf \cite{Ben09}, displayed using dashed lines. It can be seen that the agreement with the ground and first excited potential energy curve of the cation is very good. The energy gap between the ground state and the first excited state, which is important for determining the positions of the resonance states, is the same in both studies.
	Slight differences appear for the two higher excited states, b$\,^3\Pi$, c$\,^3\Delta$, due to the use of a larger basis set (aug-cc-pV5Z basis set) in \cite{Ben09}.
	
	\subsection{Scattering calculations}
	For the scattering calculations, we used the NS$^+$ MCSCF molecular orbitals and target CAS presented in section \ref{target-ss}, and the same cc-pVQZ basis set. 
	These were supplemented by continuum orbitals $u_{i,j}$ in order to represent the scattering electron.
	A truncated partial wave expansion around the centre-of-mass with $l\leq 4$ generated
	as a  GTO expansion \cite{Faure} 
	 for an R-matrix radius of 10 a$_0$. 
	Use of an orthogonalisation deletion threshold of $2\times 10^{-7}$ resulted in the removal of (10,6,6,4) orbitals from the continuum.
	The 4 (6 in C$_{2v}$ symmetry) target states shown in Figure~\ref{pes}  were used in the CC model. 

	Scattering calculations were performed as a function of bond length, $R$, 
 on a grid from 1.32 to 2.30 \AA{} with the step 0.01 \AA{} and from 2.30 to 3.0 \AA{} with a step 0.05 \AA{} giving a total of 106 points. Although the
 C$_{2v}$ point group was used for calculation,  results were extracted for $^2\Sigma^+$, $^2\Pi$, and $^2\Delta$ symmetry of the neutral system.
 
In the outer region the R-matrix was propagated \cite{Morgan84} to 150 a$_0$.
Module 	RESON \cite{Tenn84} was used to automatically detect 
 resonances and fit them to a Breit-Wigner profile to determine their position ($E_r$) and width ($\Gamma$) using grids of eigenphase sums computed for each resonance.

		\begin{figure}[t]
		  \includegraphics[width=1.15\columnwidth]{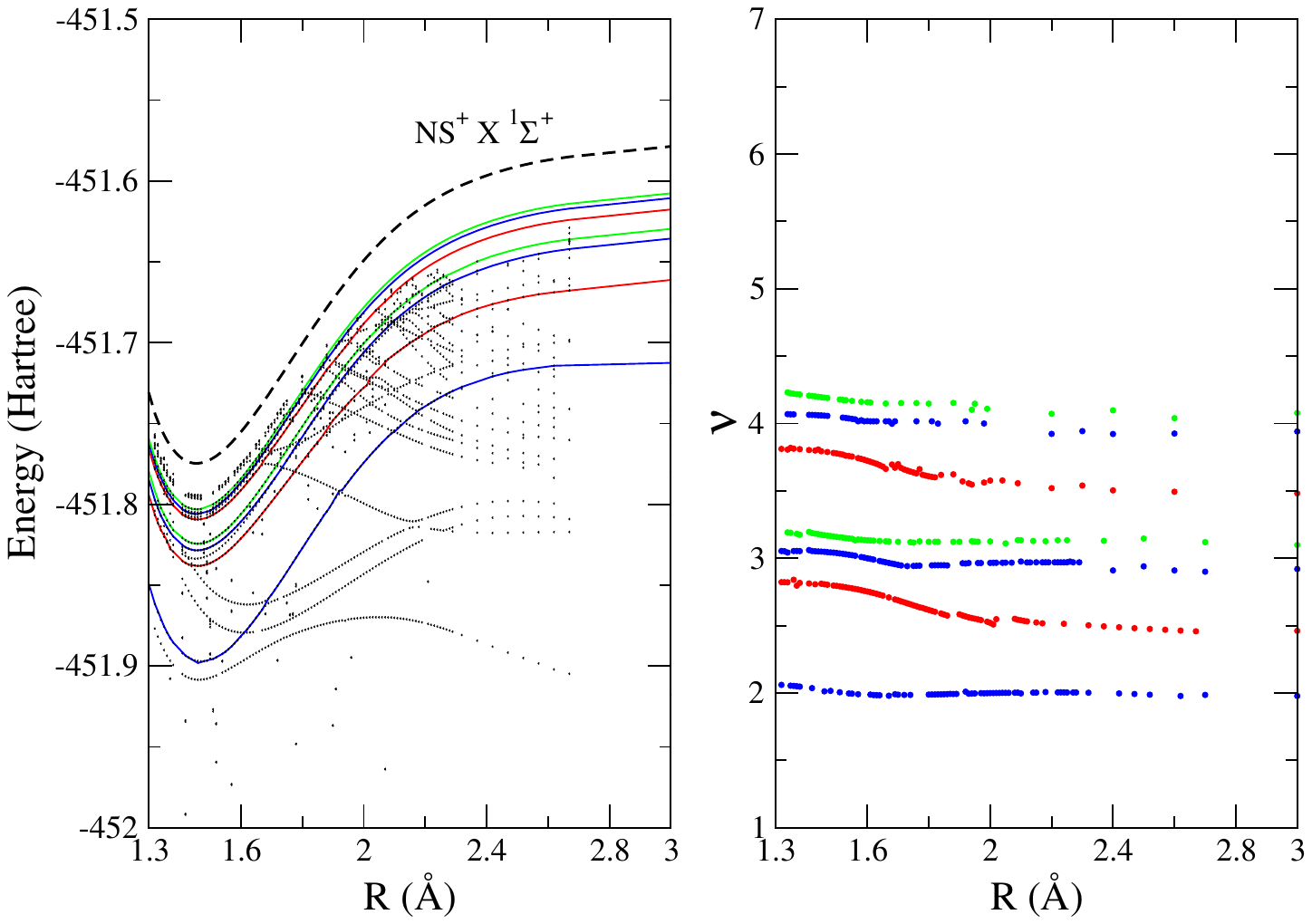}
		\caption{ NS $^2\Sigma^+$ bound states: Left panel PECs. Right panel effective principal quantum numbers. 
			Bound states belong to series that converge to $X\,^1\Sigma^+$ symmetry of ion. 
			The partial wave characterizing the Rydberg $\sigma$-wave electrons is indicated using colours: blue $s$, red $p$, green $d$.}
		\label{qn-sigma}
	\end{figure}

	\section{Results and discussion}
	In this section, we present results for the bound states and for the autoionizing resonance positions and widths of the neutral NS molecule. 
	 
	\subsection{Bound states}
	Bound electronic states of NS were found using the method of Sarpal {\it et al.} \cite{sarpal}:  R-matrices were propagated to 30 a$_0$ and wave functions computed using an improved Runge-Kutta-Nystrom algorithm \cite{zhang11}.
	Effective principal quantum numbers ($\nu$) as a function of internuclear separation were calculate from the Rydberg states assuming as threshold of the ground state of the ion. 
	
	Bound states were found for $^2\Sigma^+$, $^2\Pi$ and $^2\Delta$ total symmetries. These states
	are a mixture of Rydberg states which follow the shape of the NS$^+$ X~$^1\Sigma^+$ state and
	valence states which do not. Both sets of states are depicted in our figures. In this section we concentrate characterizing the Rydberg series but below we also consider those bound state
	curves which link with the key, dissociative resonances at large $R$.
	
\begin{figure}	
		  \includegraphics[width=1.15\columnwidth]{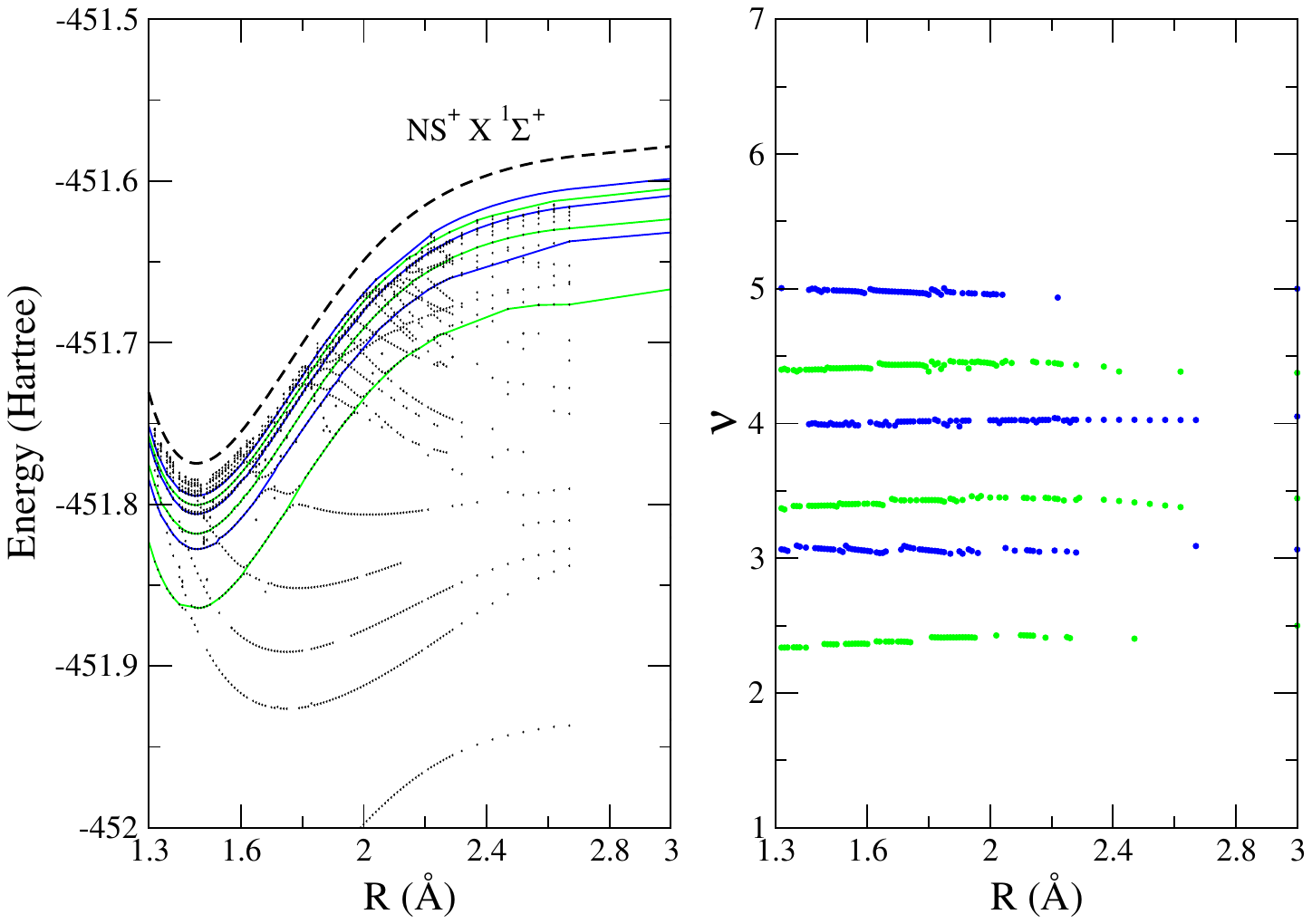}
		\caption{ NS $^2\Pi^+$, bound states: Left panel PECs. Right panel effective principal quantum numbers. 
			Bound states belong to series that converge to $X\,^1\Sigma^+$ symmetry of ion. 
			The partial wave characterizing the Rydberg $\pi$-wave electron is indicated using colours:   green $p$-wave, blue $d$-wave.
	}
		\label{qn-pi}
	\end{figure}

\begin{figure}	
		  \includegraphics[width=1.15\columnwidth]{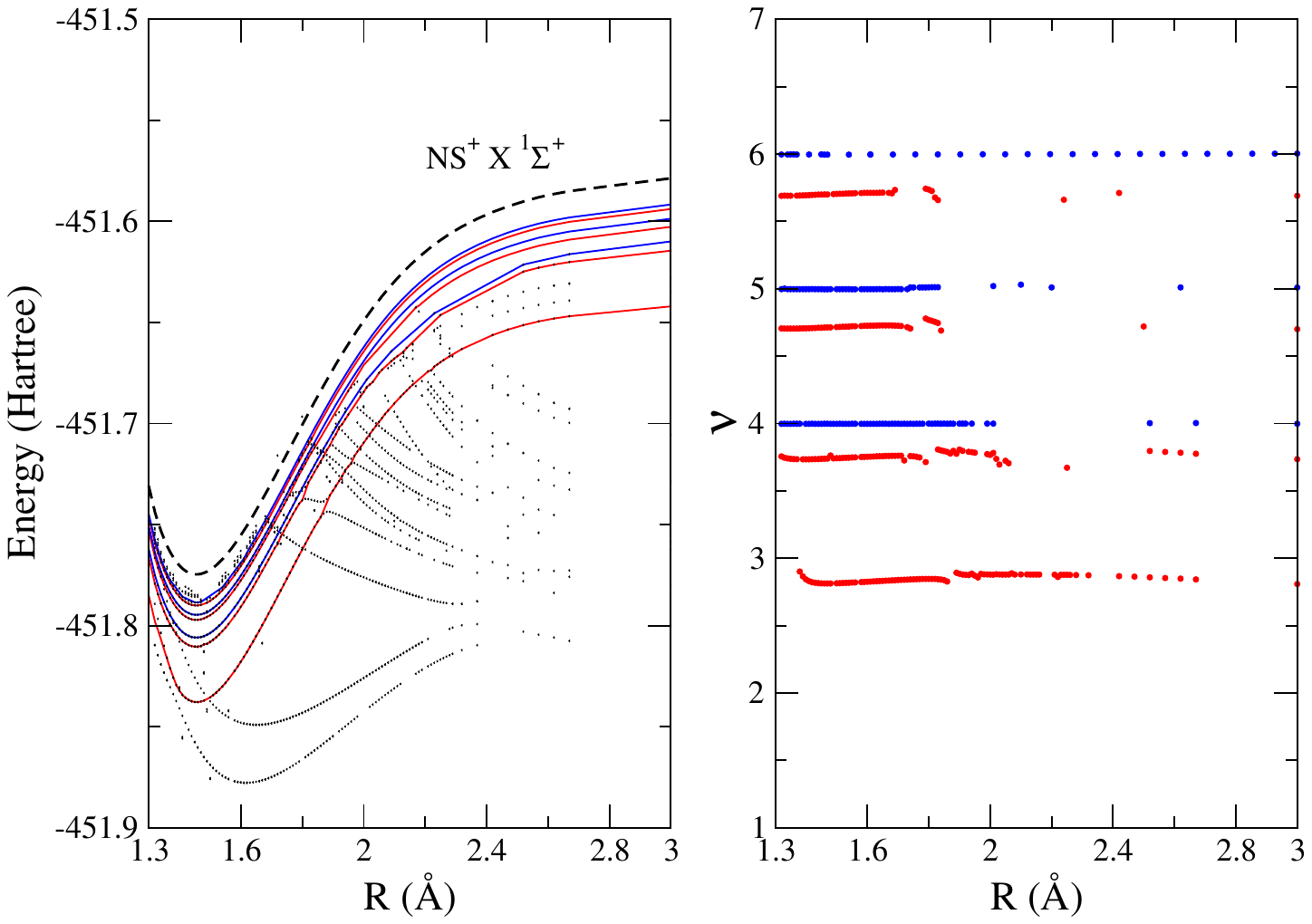}
		\caption{NS $^2\Delta^+$, bound states: Left panel PECs. Right panel effective principal quantum numbers. 
			Bound states belong to series that converge to $X\,^1\Sigma^+$ symmetry of ion. 
			The partial wave characterizing the $\delta-wave$ electron of the states is indicated using colours: red $d$-wave, blue $f$-wave. 
		}
		\label{qn-delta}
	\end{figure}

	The curves and the effective principal quantum numbers for states we assign as Rydberg-like are shown in figures \ref{qn-sigma}, \ref{qn-pi} and \ref{qn-delta}. In each figure, the uppermost dashed black curve gives the $X\,^1\Sigma^+$ ground state of NS$^+$.
	

	 If one allocates a dominant partial wave as characterizing the Rydberg electron of each states, one gets  different  Rydberg series for the various total symmetries. Note our calculations
	 neglect Rydberg series arising partial waves with $\ell > 4$; such states are expected to have quantum defects close to zero. Below we consider the major (i.e. low
	 $\ell$) partial waves for each symmetry.
	 
	 For $^2\Sigma^+$ symmetry we consider  $s$, $p$ and $d$ Rydberg electrons; for  $^2\Pi$ symmetry we  consider $p$ and $d$ electrons and for $^2\Delta$ symmetry we consider $d$ and $f$ electrons.
	 In general the effective principal quantum  numbers of the states show smooth and generally slow
	 variation with $R$. However, in places 
	 the structure of  the Rydberg series are  complicated  by the presence of 
	 so-called intruder states which arise from Rydberg series associated with excited states of the ion. Crossings by these intruder states appear as local discontinuities in the $\nu$ as
	 a function of $R$.
	
	\subsection{Resonance curves}

	 The R-matrix calculation is made in the fixed nuclei approximation and hence the PECs are adiabatic. Resonances were characterised up to approximately 4 eV above the equilibrium energy of the X$\,^1\Sigma^+$ state of ion, which restricts the internuclear distance range to  $R \approx 1.2 - 1.8$ \AA. 
	
\begin{figure}
		  \includegraphics[width=1.15\columnwidth]{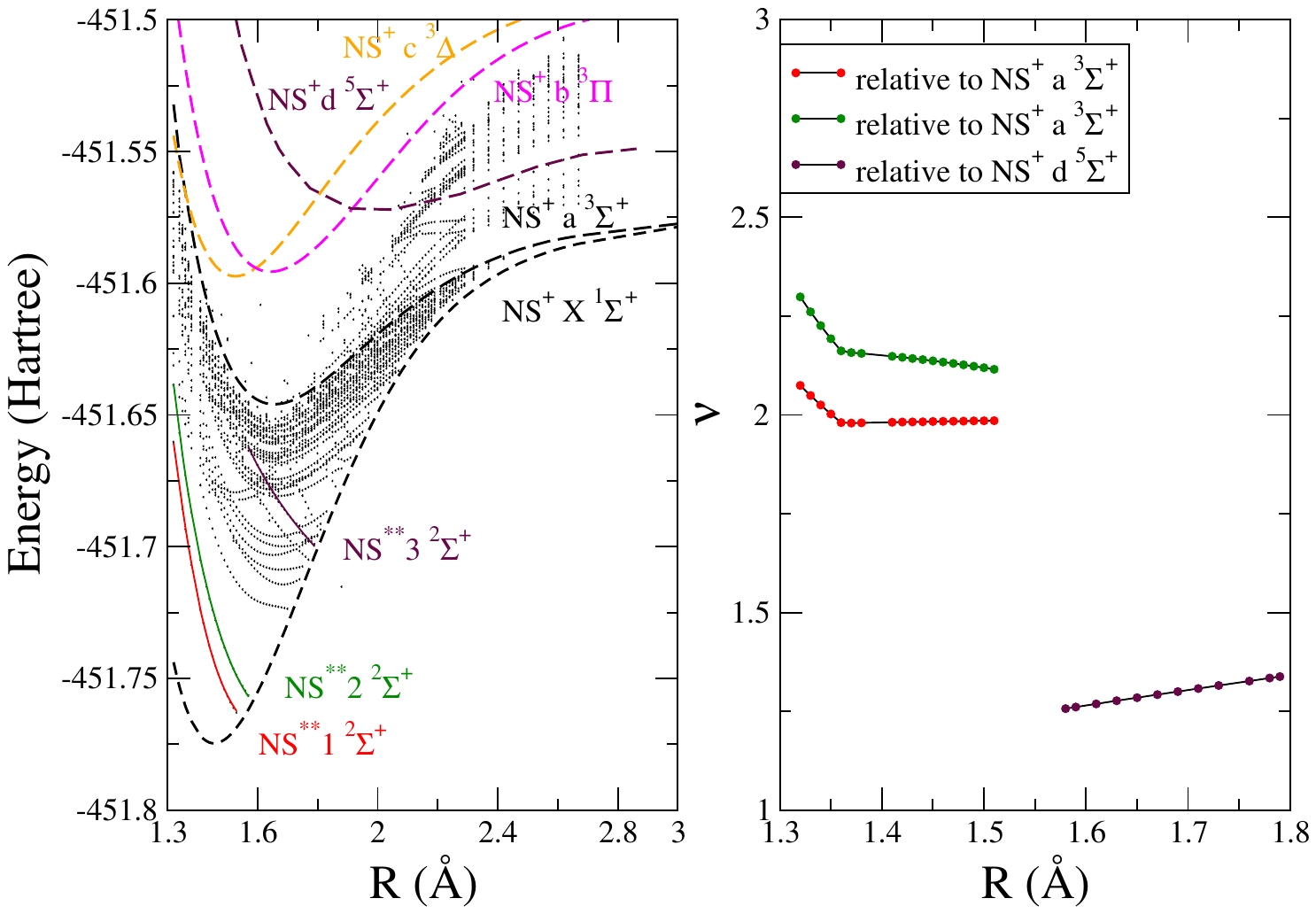}
		\caption{NS $^2\Sigma^+$ symmetry relative to  resonances curves (left panel) and effective principal quantum numbers (right panel) for some specific curves of special interest for DR.}
		\label{rez-qd-S}
	\end{figure}

\begin{figure}
		  \includegraphics[width=1.15\columnwidth]{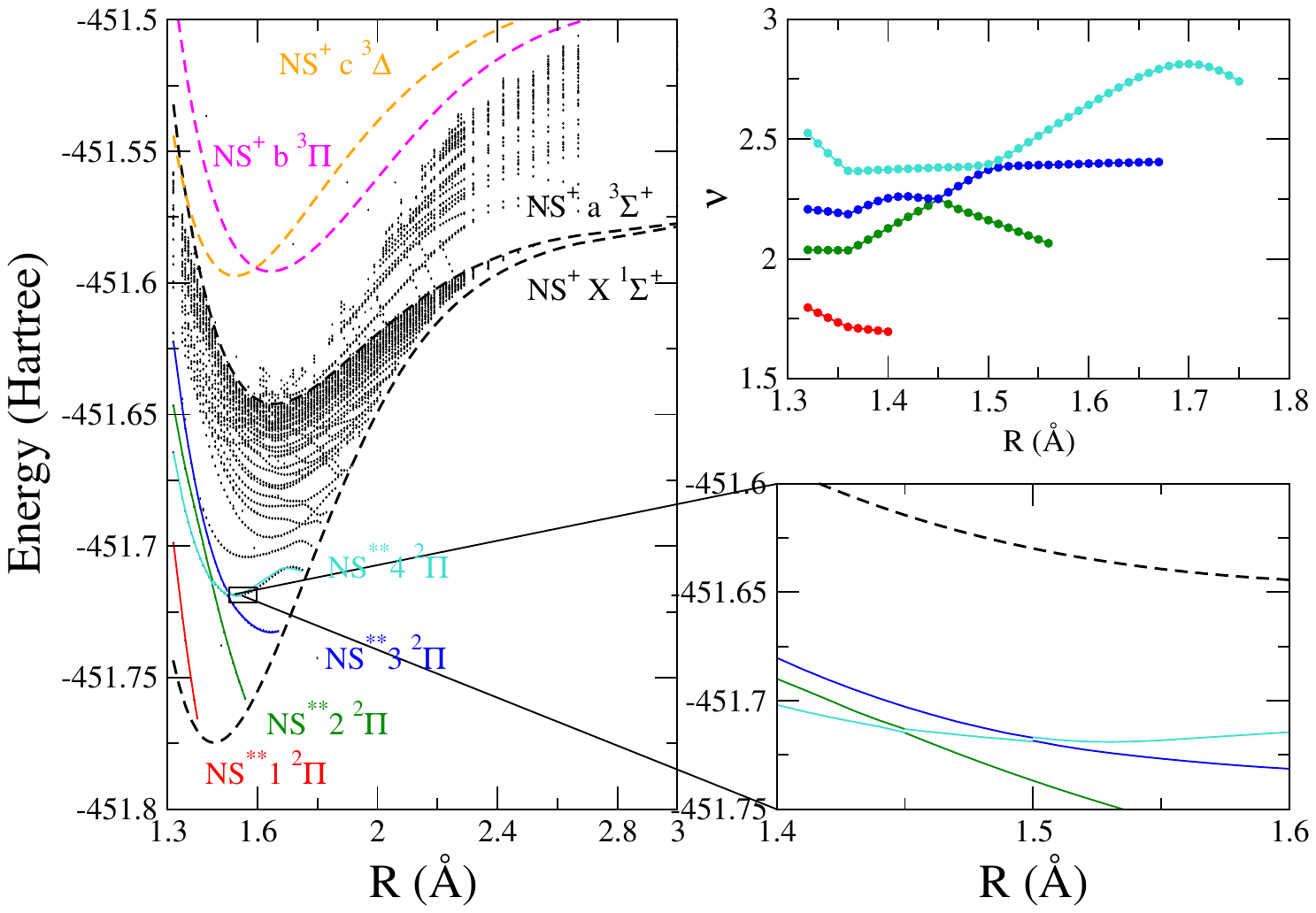}
		\caption{NS $^2\Pi$ symmetry resonances curves (left panel) and effective principal quantum numbers (right panel) for some specific curves of special interest for DR.}
		\label{rez-qd-P}
	\end{figure}

\begin{figure}
	  \includegraphics[width=1.15\columnwidth]{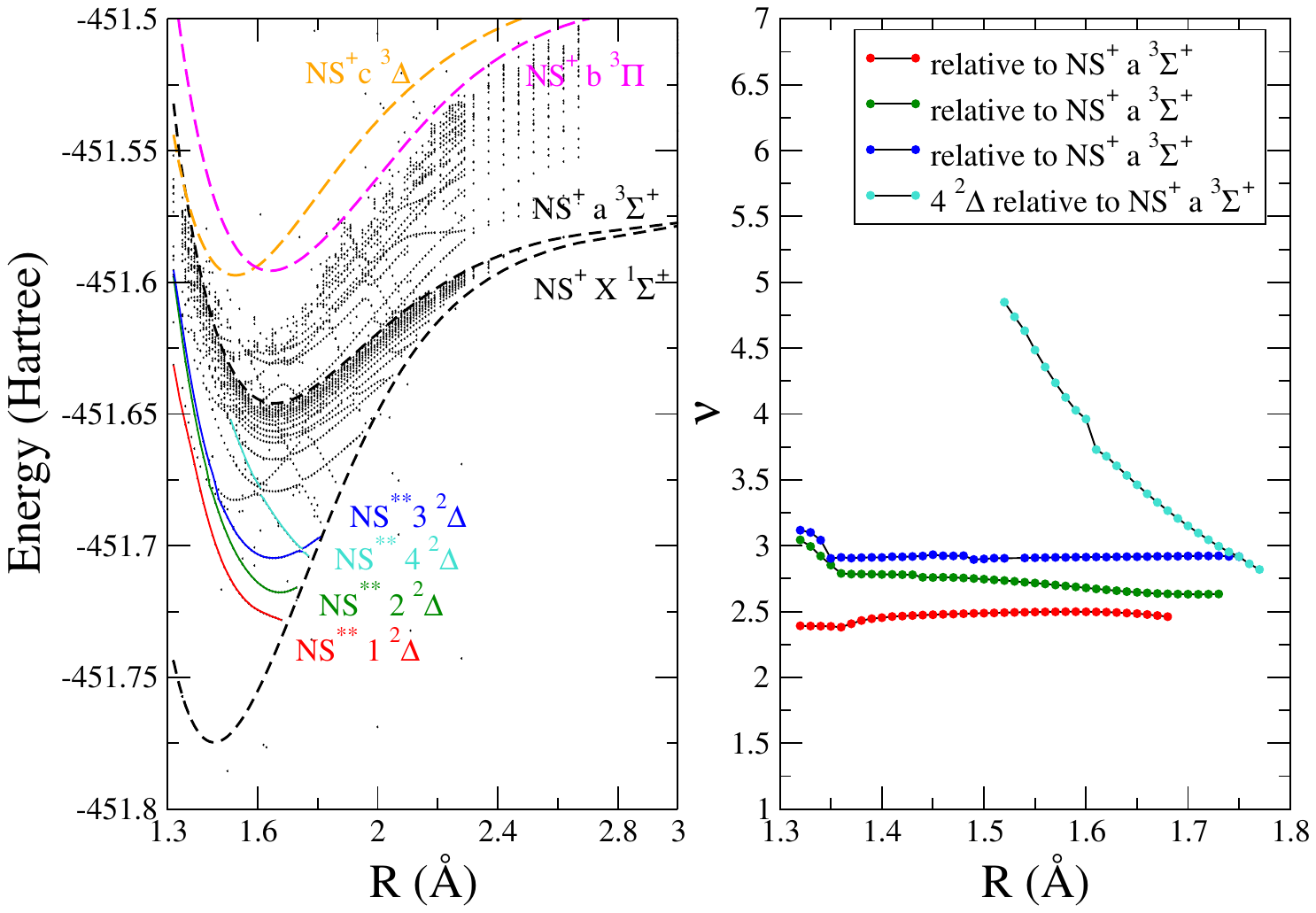}
	\caption{NS $^2\Delta$ symmetry resonances curves (left panel) and effective principal quantum numbers (right panel) for some specific curves of special interest for DR.}
	\label{rez-qd-D}
\end{figure}

	Figures \ref{rez-qd-S}, \ref{rez-qd-P}, and \ref{rez-qd-D} shows the resonances that are found with 
	emphasis on the PECs that can lead to dissociation as 
	DR occurs along these repulsive PECs.
	The right panel of each figure shows the real part of the $\nu$ for each resonance  as a function of $R$; being derived assuming that the resonance is associated with the first, a~$^3\Sigma^+$, excited state. In each figure, the curves can be matched by their colours.
	Again we concentrate on states of low $\ell$ as higher $\ell$ states have quantum defects
	close to zero and, more importantly, very weak electronic couplings meaning that in general they do not play a
	significant role in DR.
	
	Figure \ref{rez-qd-S} shows that the lowest two curves show almost constant effective principal quantum number for $R> 1.4$~\AA.  The inflection point at $R= 1.36$ \AA{} is due to the NS$^+$ c$\,^3\Delta$ curve, which for  $R < 1.36$ \AA{} crosses below the  a$\,^3\Sigma$ state. 
	The $\nu$ for the highest, black curve in Fig.~\ref{rez-qd-S} had to be given
	special consideration. It showed strong variation with $R$ when taken relative to the
a$\,^3\Sigma^+$ state. Assuming this resonance to be an intruder we plotted its effective principal quantum number relative to both the b$\,^3\Phi$ and c$\,^3\Delta$ states but again found strong	 
	variations with $R$. Finally, we took $\nu$ relative to the d$\,^5\Sigma^+$ curve, as calculated by 
	  Ben Yaghlane and Hochlaf \cite{Ben09}; this curve is the extra one shown on this figure. This showed much flatter variation although there is a pronounced slope which is possibly caused by the fact that this is not a like-for-like comparison.

	The $^2\Pi$ Rydberg states curves presented  in figure \ref{rez-qd-P} mostly showing
	little variation in $\nu$ as for $R > 1.4$~\AA.
	However, the $^2\Pi$ symmetry shows avoided crossings between the green with turquoise curves, and the turquoise and blue curves. The approximately constant behaviour of $\nu$ with $R$ is maintained with the corresponding change of colours.
	In the figure \ref{rez-qd-D} the effective principal number shows an inflexion point at $R < 1.36$ \AA{} due to the crossing of NS$^+$ c$\,^3\Delta$ curve below the a$\,^3\Sigma$ one. The effective principal number corresponding to the NS$^{**}$ 3 $^2\Delta$ state shows a strong variation with $R$ and a shift at $R = 1.6$ \AA. A comparison with a higher excited curve might show an effective quantum number with a flatter dependence on $R$.


\begin{figure}
		  \includegraphics[width=1.15\columnwidth]{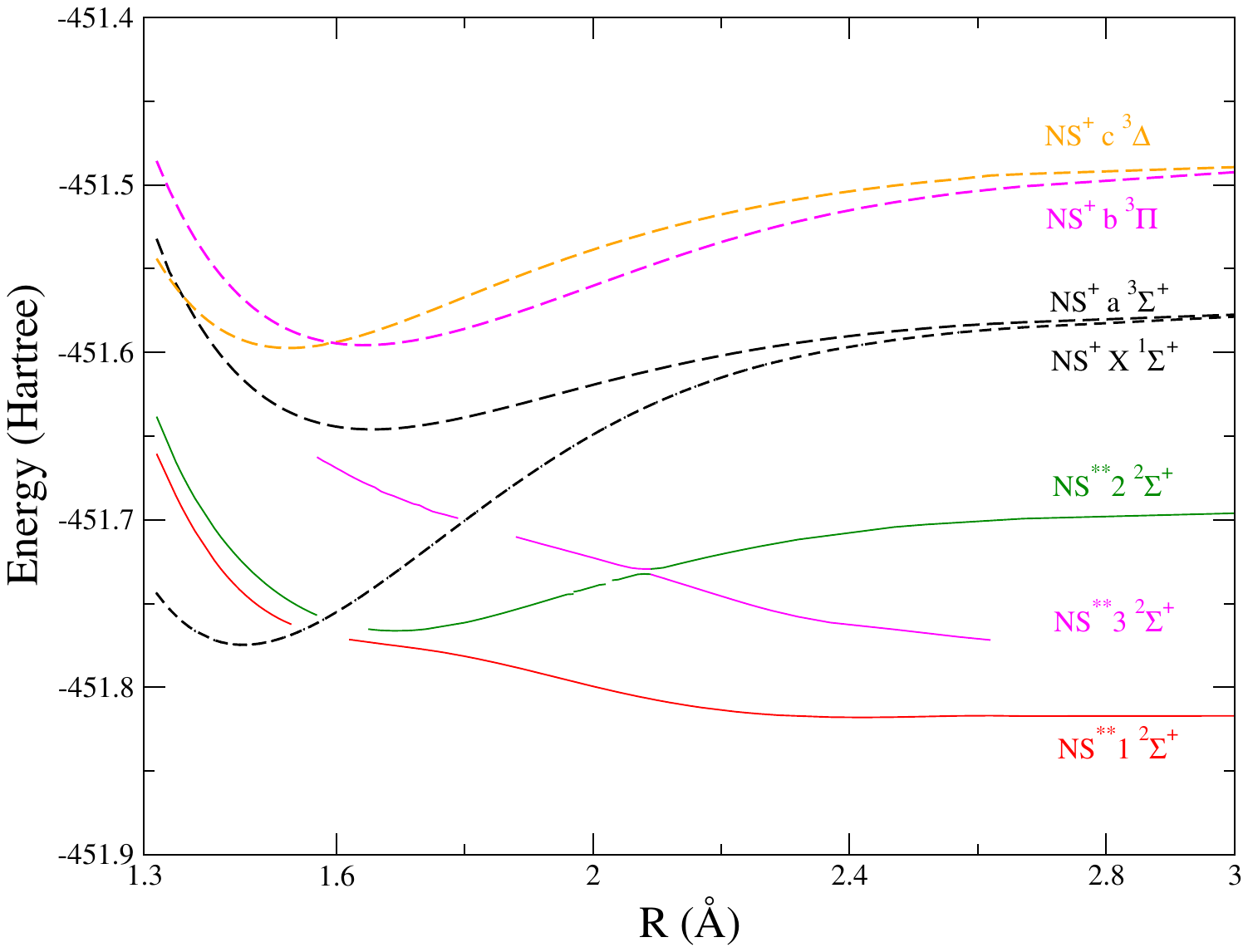}
		\caption{NS resonance and the bound states extension for the $^2\Sigma^+$ molecular symmetry as a function of internuclear distance.}
		\label{diss-A1}
	\end{figure}

\begin{figure}
		  \includegraphics[width=1.15\columnwidth]{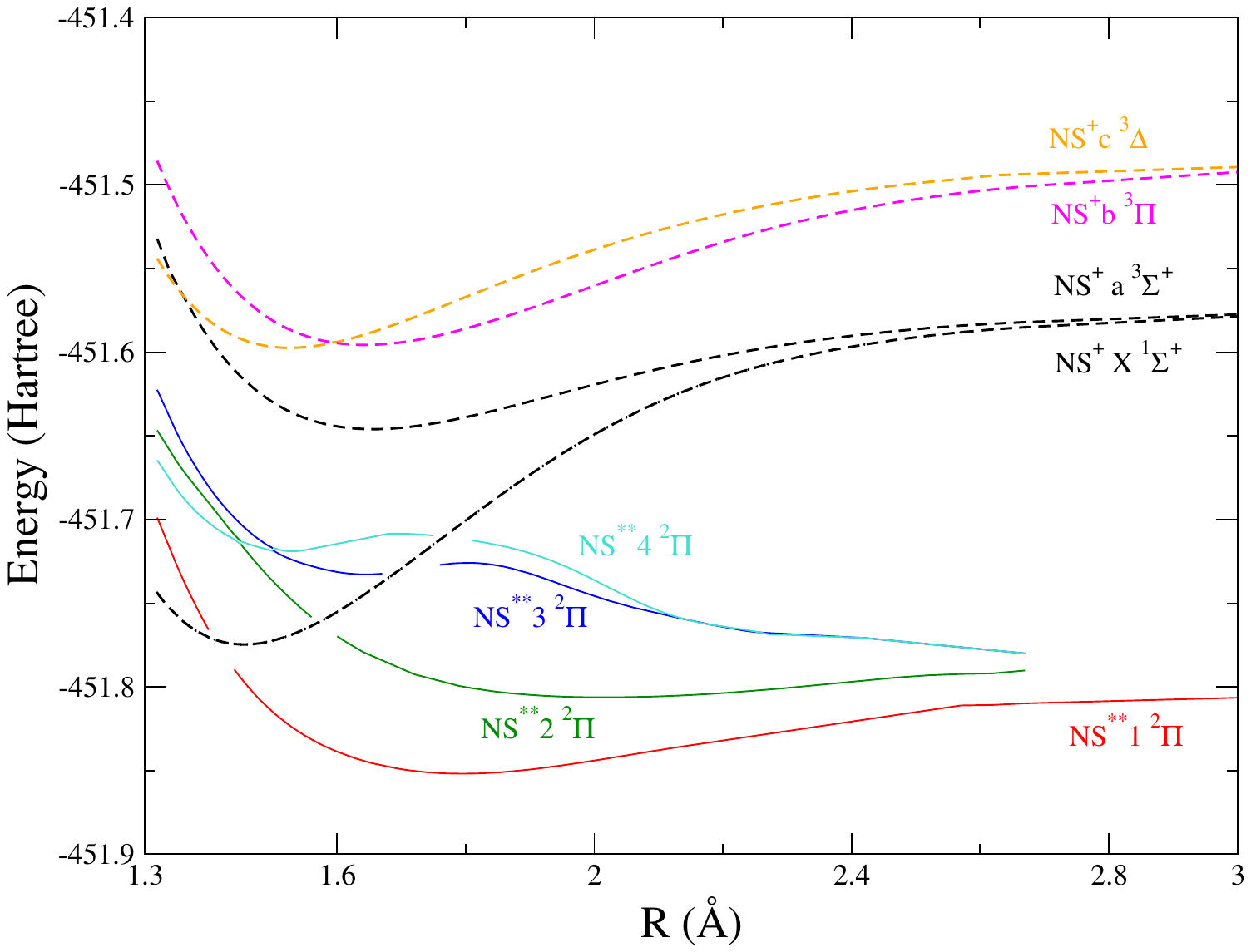}
		\caption{NS resonance and the bound states extension for the $^2\Pi$ molecular symmetry as a function of internuclear distance.}
		\label{diss-B1}
	\end{figure}

\begin{figure}
		  \includegraphics[width=1.15\columnwidth]{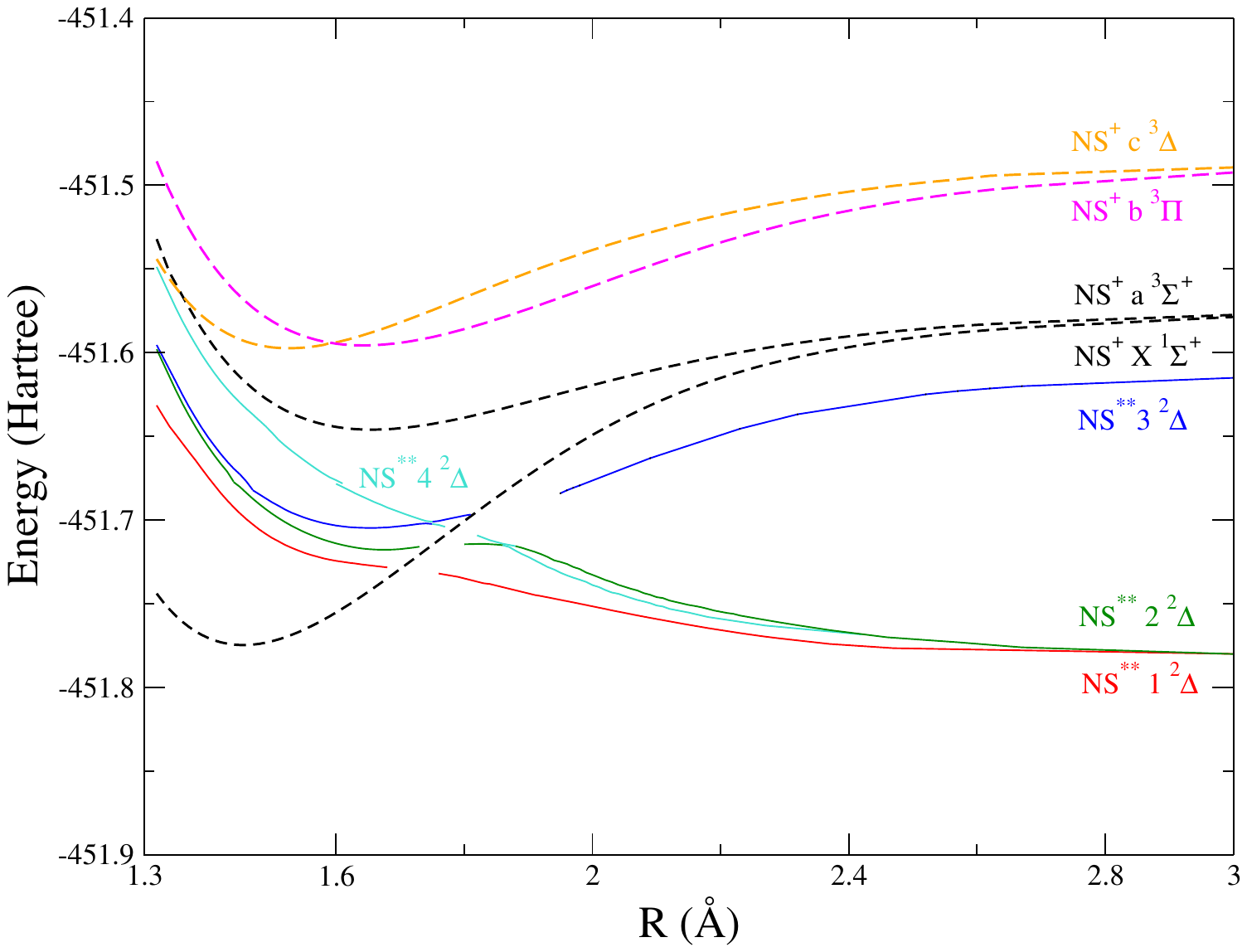}
		\caption{NS resonance and the bound states extension for the $^2\Delta$ molecular symmetry as a function of internuclear distance.}
		\label{diss-A2}
	\end{figure}

	Figure \ref{diss-A1} shows the PECs with $^2\Sigma^+$ symmetry of neutral   NS in an adiabatic representation. However, following the colours one can see the diabatic states. 
	The curves situated above the PEC of the ion ground state  are the resonances and those below it represent bound states.
	Dissociative PECs of importance for DR comprise of a resonant state in the continuum which cross the ion ground state to become bound states where they then, in the diabatic picture, pass 
	through the Rydberg series associated with the ion ground state. 
	The PECs are smooth except in a vicinity of avoided crossings. 
	Figures \ref{diss-B1} and \ref{diss-A2}  show similar curves for both $^2\Pi$ and $^2\Delta$ symmetry respectively. These curves are the ones which form the input for a model of DR.
	

	\subsection{Widths and quantum defects}

	Complex quantum defects $\mu = \alpha + i\beta$ were obtained from the fitted position and width using the formulae:
	\begin{equation}
	E_r = E_t - \frac{Ry}{\nu^2}, \;\; \Gamma = \frac{2\beta \, Ry}{\nu^3},
	\end{equation}
	where the effective principal quantum number $\nu=n-\alpha$, E$_t$ is the energy of the threshold to which the Rydberg series
	converges and $Ry$ is the Rydberg constant. 
	
\begin{figure}
		  \includegraphics[width=1.15\columnwidth]{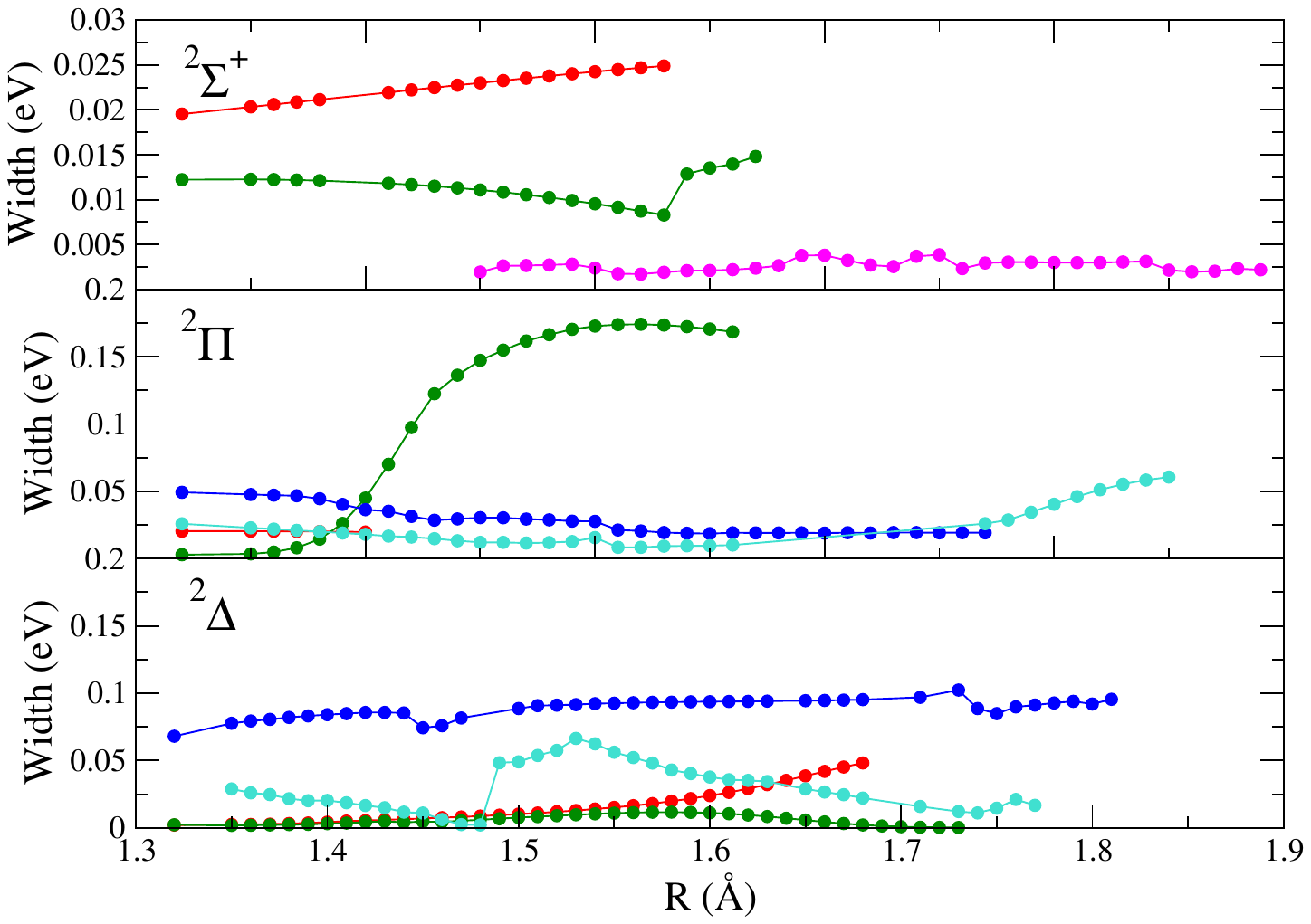}
		\caption{Resonance widths as a function of internuclear distance in agreement with the resonances from figures \ref{diss-A1}, \ref{diss-B1}, and \ref{diss-A2}. The corresponding symmetry of the resonant states is indicated in the ﬁgure.}
		\label{widths}
	\end{figure}

	The widths are presented in figure \ref{widths} as functions of $R$; the symmetry is marked on each graph. One may see how the behaviour of widths change when avoided crossings occur. 
	In general, we expect the width of a resonance to decrease as the number of open channels decreases which occurs when a resonances passes through a state of the ion.
Note that the  widths should vanish once the resonance states cross the ion ground state and become bound.
	
 	\section{Conclusions}
	
	
	Resonance positions and auto-ionization widths were calculated for e-NS$^+$ system depending on the internuclear separation. 
	The Rydberg series that converge to ion ground state were computed together with the corresponding effective principal quantum numbers. 
	The use of a dense grid produces numerous resonances and bound states in great details, which facilitate the identification of dissociative states and singly excited Rydberg manifolds.
	 Figures  \ref{diss-A1}, \ref{diss-B1},
	and \ref{diss-A2} summarize our results and provide the curves and electronic couplings which will provide the
	input for future nuclear dynamics (dissociative recombination and ro-vibrational transitions) studies.
	To our knowledge, this is the first time when relevant molecular data sets were calculated to study electron induced reactive elementary processes in NS$^+$. 
	The data produced in this work will be used for calculating dissociative recombination and ro-vibrational transition cross sections and rate coefficients relevant to the astrochemistry of sulphur and nitrogen, using stepwise Multichannel Quantum Defect Theory, that has proven its power for many diatomic and polyatomic molecular systems \cite{Little14,Niyonzima17,Niyonzima18,Pop21,Epee22,slava18}.
	
	\section*{Acknowledgements}
	We are grateful to Kalyan Chakrabarti for helpful discussions.
	This article is based upon work performed as part of COST Action: CM1401 - Our Astro-Chemical History, CA17126 - TUMIEE, CA18212 - MD-GAS.
	The authors acknowledge support from La R\'egion Normandie, FEDER, and LabEx EMC3 via the projects PTOLEMEE, Bioengine COMUE Normandie Universit\'e, the Institute for Energy, Propulsion and Environment (FR-IEPE), and to Agence Nationale de la Recherche (ANR) via the project MONA. This work was supported by the Programme National “Physique et Chimie du Milieu Interstellaire” (PCMI) of CNRS/INSU with INC/INP co-funded by CEA and CNES. JZsM thanks the financial support of the National Research, Development and Innovation Fund of Hungary, under the K 18 and FK 19 funding schemes with project numbers K 128621 and FK 132989.
	JZsM is greateful for the hospitality of West University Timi\c soara within the Visiting@WUT grant programme. JT and JZsM are indebted to the Distinguished Guest Scientists Fellowship Programme-2022 of the Hungarian Academy of Sciences.  TM was funded by EPSRC (Grant No. EP/M507970/1). 
\section*{Data availability} Upon a reasonable request, the data supporting this article will be provided by the corresponding author.

\end{document}